\documentclass[%
 reprint,
 aps,
 nofootinbib,
]{revtex4-2}

\preprint{APS/123-QED}

\usepackage{mathrsfs,mathpazo,amsmath,mathtools,physics}

\usepackage{graphicx}
\usepackage{dcolumn}
\usepackage{hyperref}
\usepackage{xcolor}
\usepackage{ulem}
\usepackage{tikz}
\usepackage{listings}
\usepackage{lipsum, babel}
\usepackage{multirow}

\usepackage[makeroom]{cancel}
\usepackage{float}
\usepackage{pgf}

\tikzset{
    level/.style = {
        ultra thick,
        black,
    },
    connect/.style = {
        dashed,
        red
    }
}
\usetikzlibrary{backgrounds,calc}
\usetikzlibrary{shapes.misc}
\usepackage{multirow}
			
\usepackage[font=small,labelfont=bf,justification=justified]{subcaption}
\usepackage[font=small,labelfont=bf,justification=justified]{caption}

\allowdisplaybreaks
\makeatletter
\newcommand*\rel@kern[1]{\kern#1\dimexpr\macc@kerna}
\newcommand*\widebar[1]{%
  \begingroup
  \def\mathaccent##1##2{%
    \rel@kern{0.8}%
    \overline{\rel@kern{-0.8}\macc@nucleus\rel@kern{0.2}}%
    \rel@kern{-0.2}%
  }%
  \macc@depth\@ne
  \let\math@bgroup\@empty \let\math@egroup\macc@set@skewchar
  \mathsurround\z@ \frozen@everymath{\mathgroup\macc@group\relax}%
  \macc@set@skewchar\relax
  \let\mathaccentV\macc@nested@a
  \macc@nested@a\relax111{#1}%
  \endgroup
}

\definecolor{lime}{HTML}{A6CE39}
\DeclareRobustCommand{\orcidicon}{
	\begin{tikzpicture}
	\draw[lime, fill=lime] (0,0) 
	circle [radius=0.16] 
	node[white] {{\fontfamily{qag}\selectfont \tiny ID}};
	\draw[white, fill=white] (-0.0625,0.095) 
	circle [radius=0.007];
	\end{tikzpicture}
	\hspace{-2mm}
}
\foreach \x in {A, ..., Z}{\expandafter\xdef\csname orcid\x\endcsname{\noexpand\href{https://orcid.org/\csname orcidauthor\x\endcsname}{\noexpand\orcidicon}}}
\begin{document}

\preprint{APS/123-QED}

\title{Quasiuniversal relations in the context of future neutron star detections}

\author{Lami Suleiman\orcidA{}}
\email{lsuleiman@fullerton.edu}
 \affiliation{Nicholas and Lee Begovich Center for Gravitational Wave Physics and Astronomy, California State University Fullerton, Fullerton, California 92831, USA}
\def\ls{\textcolor{purple}}

\author{Jocelyn Read\orcidB{}}
 \affiliation{Nicholas and Lee Begovich Center for Gravitational Wave Physics and Astronomy, California State University Fullerton, Fullerton, California 92831, USA}
\def\jr{\textcolor{blue}}
\date{\today}

\begin{abstract}
% \begin{description}

%\item[Background] 
The equation of state dependence of a neutron star's astrophysical features is key to our understanding of isospin asymmetric and dense matter. There exists a series of almost equation of state independent relations reported in the literature, called quasiuniversal relations, that are used to determine neutron star radii and moments of inertia from x-ray and gravitational wave signals.
%\item[Purpose]
Using sets of equations of state constrained by multimessenger astronomy measurements and nuclear-physics theory, we discuss quasiuniversal relations in the context of future gravitational wave detectors Cosmic Explorer and Einstein Telescope, and the Spectroscopic Time-Resolving Observatory for Broadband Energy X-rays.
%\item[Methods]
We focus on relations that involve the moment of inertia $I$, the tidal deformability $\Lambda$, and the compactness $C$: $C(\Lambda)$, $I(\Lambda)$ and $I(C)$. The quasiuniversal fits and their associated errors are constructed with three different microphysics approaches which include state of the art nuclear physics theory and astrophysical constraints. Gravitational-wave and x-ray signals are simulated with the sensitivity of the next generation of detectors. Equation of state inference on those simulated signals is compared to determine if it will offer a better precision on the extraction of a neutron star's macroscopic parameters than quasiuniversal relations.
%\item[Results]
We confirm that the relation $I(\Lambda)$ offers a more pronounced universality than relations involving the compactness regardless of the equation of state set. We show that detections with the third generation of gravitational wave detectors and future x-ray detectors will be sensitive to the fit error marginalization technique. We also find that the sensitivity of those detectors will be sufficient in that using full equation of state distributions leads to significantly better precision on extracted parameters than quasiuniversal relations. 
%\item[Conclusion] 
%The equation of state independence of quasiuniversal relations that involve the compactness have to be carefully contrasted with the sensitivity of the detector. 
We also note that nuclear physics theory offers a more pronounced equation of state invariance of quasiuniversal relations than current astrophysical constraints. 
% \end{description}
\end{abstract}

\keywords{}
\maketitle

%-------------------------------------------------------------------------------------%
%-----------------------------------  Sec. 1  -------------------------------------%
%-------------------------------------------------------------------------------------%

\section{Introduction}

As the densest stars in the Universe, neutron stars are particularly well suited to investigate ultradense matter. Our understanding of the innermost layers of neutron stars remains uncertain due to limitations of nuclear physics laboratories to reach equivalent regimes of temperature and density. Neutron star astrophysical features strongly depend on the equation of state of ultradense matter, thus offering the opportunity to probe the neutron star interior with multimessenger astronomy. On the other hand, a series of relations between various neutron star observables, referred to as quasi (or almost) universal relations, were empirically found to depend weakly on the exact equation of state \cite{Yagi2013}. 

While the perfect universality (as per the no-hair theorem) of isolated and stationary black holes in the gravitational theory of general relativity is, in principle, not applicable to compact stars, their external gravitational field presents features that are almost independent of the neutron star's interior. This (almost) universality applies to nonmagnetized neutron stars on a static metric in general relativity and holds for magnetized (e.g., \citep{Haskell2014}) and spinning neutron stars (e.g., \citep{Chakrabarti2014, Stein2014, Pappas2014}), as well as in modified gravity theories (e.g., \citep{Doneva2015}). The underlying physics of this universality has been connected to approximate no-hair relations, where high order multiple moments of compact stars are approximately determined by low order multipole moments (e.g., \cite{Chatziioannou2014,Yagi2014}) , and to the self-similarity of isodensities in compact stars (e.g., \citep{Yagi2014b}). For a detailed introduction and history of universality in compact stars, as well as the derivation of the approximate no-hair relations, details on the I-Love-Q relations and also on the black-hole limit of this almost universality, we refer to the extensive review and work of Ref.~\cite{Yagi2017} and references therein. 

Quasiuniversal relations between macroscopic properties of neutron stars are well suited to extract one parameter from the measurement of others. The era of multimessenger astronomy has allowed for the detection of various astrophysical neutron star features and is expected to provide increasingly more precise observational data in the future. The measurement of post-Keplerian parameters in binary systems via pulsar timing has provided the most precise measurements of neutron star's gravitational mass $M$, see e.g., Refs~\cite{ Lattimer2012,Antoniadis2013,Freire,Ozel2016,Alsing2018}. Various wavelengths of the electromagnetic spectrum (radio, x-ray and optics), as well as gravitational wave signals from binary neutron star mergers, have provided a large number of mass measurements\footnote{For a list of neutron star masses measurements with reported precision, see \url{https://compose.obspm.fr/resources}.}. The Neutron star Interior Composition ExplorER (NICER) telescope, which relies on the effects of general relativity on the detection of hot spots on the surface of a rotating neutron star, has provided the simultaneous measurement of both the gravitational mass and the radius $R$ for J0030$+$0451  \citep{miller2019, riley2019} and of $R$ for J0740$+$6620 \citep{miller2021, riley2021} with an independent mass measurement based on radio pulsar timing \cite{Fonseca2021}. 
The future Spectroscopic Time-Resolving Observatory for Broadband Energy x-rays (STROBE-X) \citep{Ray2019} is expected to offer measurements of mass-radius contours two to three times tighter. The tidal deformability $\Lambda$ of a neutron star was constrained for the first time from the gravitational wave measurement of the double neutron star binary merger source GW170817 \citep{Abbott2017}. The significant increase in sensitivity of the next generation of gravitational wave detectors, such as the Cosmic Explorer (CE) \cite{CE2021} and the Einstein Telescope (ET) \cite{ET2020, Punturo2010, Branchesi-2023}, is expected to increase the number and precision of such observations by orders of magnitude. Double pulsar binaries are the most promising systems to detect the star's moment of inertia $I$; the famous double pulsar PSR J0737$-$3039 for which more than 15\;years of data was gathered \citep{Kramer2021}, has not yet permitted the direct measurement of the moment of inertia. However, Ref.~\cite{Landry2018} has been able to use a quasiuniversal relation that involves $I$, $M$ and $\Lambda$ to extract constraints on the moment of inertia of J0737$-$3039A.

In this paper, we assess the quasiuniversality of three relations and discuss their usefulness in the context of next generation of detectors. We compare the precision achieved on the radius and moment of inertia of a source when using quasiuniversal relations to that derived from full equation of state inference. In Sec.~\ref{sec:methods}, the quasiuniversal relations used in this paper and the equation of state sets on which they are based are discussed. We then give details on quasiuniversal relations fits and how to introduce a fit error based on the set of equations of state considered. In Sec.~\ref{sec:results}, we compare the quasiuniversal relations designed with different equation of state sets. We simulate next-generation detections, and demonstrate the use of quasiuniversal relations to extract parameters that are not directly measured. We then discuss the impact of different marginalizations of the fit error on the extraction of neutron star's macroscopic parameters. Finally, we use equation of state inference from the same simulated detections, and compare to the accuracy of parameter extraction with quasiuniversal relations. In the Appendix, we present the mass, radius, tidal deformability and moment of inertia modeling, as well as parameters for the quasiuniversal relation fits. 

%-------------------------------------------------------------------------------------%
%-----------------------------------  Sec. 2  -------------------------------------%
%-------------------------------------------------------------------------------------%
\section{Methods}\label{sec:methods}

In this paper, we explore astrophysical features of a neutron star within the theory of general relativity.  The mass $M$ and radius $R$ are found using the Tolman-Oppenheimer-Volkov (TOV) \citep{Tolman1939, Oppenheimer1939} differential equations closed by an equation of state (EoS). The moment of inertia $I$ is determined in the slow-rotation approximation of Ref.~\cite{Hartle1967} and the tidal deformability follows the quadrupole perturbation derivation of Ref.~\cite{Hinderer2008} derivation. For details on the modeling of macroscopic parameters, see Appendix~\ref{sec:appendix,modeling}.

%-------------------------------------------------------------------------------------%
%-------------------------------------------------------------------------------------%
\subsection{Quasiuniversal relations}

%-------------------------------------------------------------------------------------%
\subsubsection{$C(\Lambda)$, $I(\Lambda)$, and $I(C)$ }\label{sec:ur}

In this paper, we study the following quasiuniversal relations:
\begin{enumerate}
    \item The relation between the compactness and the dimensionless tidal deformability $C(\Lambda)$, with ${C=GM/(Rc^2)}$. In the assumption that this relation is equation of state independent, it has been used to extract the radius of a neutron star from the gravitational wave measurement of its mass and tidal deformability, see e.g., Ref.~\cite{Abbott2018}. 
    \item The relation between the dimensionless moment of inertia $\bar{I}=Ic^4/(G^2 M^3)$ and the dimensionless tidal deformability $\bar{I}(\Lambda)$. In the assumption that this relation is equation of state independent, it can be used to extract the moment of inertia from the gravitational wave measurement of the mass and tidal deformability, see e.g., Ref.~\cite{Landry2018}.
    \item The relation between the dimensionless moment of inertia and the compactness $\bar{I}(C)$. In the assumption that this relation is equation of state independent, it can be used to extract the moment of inertia from the simultaneous measurement of the mass and radius by telescopes such as NICER or STROBE-X, see e.g., Ref.~\cite{Silva2021}. 
\end{enumerate}

Not all of the above mentioned relations are equally equation of state independence (see Ref.~\cite{Yagi2017,Legred2023}), but they are all described in the literature as quasiuniversal. It is also possible to parametrize them using what is referred to as fits, see, e.g., Refs.~\cite{Maselli13, Yagi2013, Yagi2017, Godzieba2021, Breu16,Yagi2016, Carson2019} or Sec.~\ref{sec:paramExtract/fits} of the present paper. Universality suggests that the relations studied in this paper should relate astrophysical features of neutron stars for any description of the neutron star interior, i.e. for any EoS. It should then be possible to use widely varying EoSs (only causal and thermodynamically consistent) and still retain the quasiuniversality in the relations. Such relations would not change due a decreased EoS variability related to our increasing knowledge on the behavior of ultradense matter, or our bias in considering certain EoSs. However, in practice, introduction of astrophysical and nuclear constraints on the EoS lead to updated fits of the quasiuniversal relations. This reflects a observation-driven narrowing of the EoS range rather than full universality. Observations can include both astrophysical data and nuclear physics data; see, e.g., Refs.~\cite{Carson2019, Godzieba2021}.

%-------------------------------------------------------------------------------------%
%-------------------------------------------------------------------------------------%
\subsection{Equations of state sets}\label{sec:eos}

In this paper, we compare quasiuniversal relations established from several observation-informed EoS sets.
%-------------------------------------------------------------------------------------%
\subsubsection{Nuclear physics based set}

The first set comprises 61 EoSs established from full nuclear physics calculations. The EoSs are gathered from Ref.~\cite{Suleiman2022} to which we add seven EoSs presented in Ref.~\cite{Pereira2016}. They describe cold and catalyzed $\beta$-equilibrated matter as such thermodynamic conditions are relevant for isolated "adult" (not proto) neutron stars and for the inspiral phase of a neutron star merger\footnote{We acknowledge that a global thermodynamic equilibrium (catalyzed) neutron star crust may not be appropriate to describe some of the potential STROBE-X sources located in accreting binaries, but nevertheless we reasonably neglect the impact of an accreted crust on the modeling of macroscopic parameters.}. 

The EoSs of this set were computed with two different nuclear physics approaches: relativistic mean field theory (x39) and Skyrme force energy density functionals (x22). Among the relativistic mean field models are included ten EoSs with a hyperonic core (for a recent review on hyperonic compact stars, see Ref.~\cite{Sedrakian2023}) and 14 hybrid models with a core quark phase transition (see, e.g., Refs.~\cite{Baym2018, Annala2020}); the rest have nucleonic cores. 

All EoSs of this set follow thermodynamic consistency; note that the Skyrme density functional is based on a nonrelativistic approach that does not guarantee sound speed causality. They are unified, meaning that the crust (low density) and the core (high density) have been calculated with the same nuclear physics model; we refer to Ref.~\cite{Suleiman2021} for a discussion on the role of nonunified EoSs in neutron star modeling. A large majority of them also permit the direct Urca process, a neutrino emission reaction which x-ray observations indicate exists in neutron stars \citep{Beznogov2015}. They all meet at least the mass constraint imposed by the 1$\sigma$ pulsar timing measurement of the millisecond pulsar J0740$+$6620 with a mass of $2.08 \pm 0.07$\;M$_{\odot}$ \citep{Fonseca2021}\footnote{In Ref.~\cite{Suleiman2022}, the mass constraint follows J1614$-$2230 measured at $1.908\pm0.016$\;M$_{\odot}$ \citep{Arzoumanian_2018}, which is why in our set we did not include H3, hyperonic DD2 and FSU2H, BSk19, KDE0v1, SKOp and BCPM}. Overall, this set comprises various core compositions and nuclear approaches, with reasonable microphysics parameters when compared to modern nuclear physics laboratory (see Ref.~\cite{Suleiman2021}) and astrophysics measurements, while keeping a certain variability in their nuclear physics features.

%-------------------------------------------------------------------------------------%
\subsubsection{Agnostic sets}

The two other sets are based on an agnostic approach, that is to say that the EoSs are not constructed with specific nuclear physics calculations. The point of such sets is to explore the parameter space of pressure and density and to go beyond our usual EoS constructions. In the following, we discuss two agnostic sets: one parametric which we will refer to as the "metamodel" set, and one nonparametric, later on referred to as the "Gaussian process" set. 

The first of the agnostic sets is based on Ref.~\cite{Davis2023}. To construct one EoS, the nuclear empirical parameters that are the saturation density, the energy of symmetric matter at saturation density, the symmetry energy, the isoscalar and isovector incompressibility, skewness and kurtosis, the nucleon effective mass and the effective mass isosplit are randomly thrown in intervals determined by nuclear physics laboratory experiments. Taking advantage of chiral-effective-field theory calculations presented in Ref.~\cite{Drischler2016} for pure neutron matter, the $\beta$-equilibrated EoSs can be constrained. For a given collection of nuclear empirical parameters, the low density part of the EoS is reconstructed according to the metamodeling approach discussed in Ref.~\cite{Margueron2018}, with a compressible liquid drop model for the inhomogeneous crust. The high density part is constructed with five polytropes for which the adiabatic index and the polytropic constant are randomly thrown. The metamodel set, respects thermodynamic consistency, sound speed causality, and provides EoSs in accordance with nuclear physics laboratory experiments while keeping the freedom permitted by the unknown core behavior and the error bars of nuclear experiments at low density. It is a parametric approach, that is to say it follows a specific functional, which implements bias. This set is constrained by the pulsar timing mass measurement of the millisecond pulsar J0740$+$6620; it is composed of $5\times 10^4$\;EoSs, and is denoted MM+$\chi$+PSR.

The second agnostic set is publicly available and constructed with the Gaussian process approach discussed in Ref.~\cite{Essick2020}. Contrary to a parametric approach, the Gaussian process EoSs are not bound by the functional chosen to parametrize and therefore avoids the sort of bias inherent to parametrized constructions \citep{Greif2019,Legred2022}, e.g., the discontinuous sound speed of piecewise polytropes in the metamodel set. The Gaussian process set at high density is trained on fifty nuclear physics based EoSs; the selection of EoSs constitutes a bias, but the training is believed to be sufficiently loose that this bias is tamed. The low density part of the EoS is conditioned on three crust EoSs. Contrary to the metamodel, this set does not enforce nuclear physics constraints on the crust, a connected core-crust transition, or $\chi$EFT constraints. It follows thermodynamic consistency and sound speed causality. This set of EoSs is constrained by the mass of J0740$+$6620, NICER measurements, and by the tidal deformability measurement of GW170817, we refer to it as GP+astro.

The relation between the pressure $P$ and the baryonic density $n_B$ for the two agnostic sets is presented in Fig.~\ref{fig:setPnb} in Appendix~\ref{app:EoSSets}.
%-------------------------------------------------------------------------------------%
%-------------------------------------------------------------------------------------%
\subsection{Parameter extraction}

%-------------------------------------------------------------------------------------%
\subsubsection{Parameter extraction from future astrophysical detections} \label{sec:UR/sources}

We refer to parameter extraction as determining the value of an astrophysical parameter from the measurement of one or several others. To do so, one can either use: 
\begin{itemize}
    \item Quasiuniversal relation fits: the measurement of two neutron star parameters lets us extract the measurement of a third directly from the relation.
    \item EoS inference: the detection of two neutron star parameters is used to additionally constrain the set of EoSs using Bayesian inference, and the constrained EoSs are used to compute the third parameter based on the measured values. 
\end{itemize}
For example, using a mass and tidal deformability measurement,  we can estimate the radius from the quasiuniversal relation $C(\Lambda)$, or we can use hierarchical inference to update the set of possible EoS, and use the resulting set to find the radius from the mass measurement.

The relations presented in Sec.~\ref{sec:ur} are preferred in parameter extraction because of their simplicity of use and the idea that we can count on the universality to retain a tight error bar on the extracted parameter. In previous usage, quasiuniversal relations were broadly equivalent in parameter extraction when compared to direct EoS inference (for example, in GW170817 Ref.~\cite{Abbott2018}). But in the context of increasing detector precision, we have to assess whether the detector precision overcomes the EoS dependence of the relations discussed in this paper. In other words, we study if quasiuniversal relations conditioned on state of the art astrophysics and nuclear physics will be sufficient for a precise parameter extraction in the context of current and future detectors or if EoS-inference with next-generation astrophysical measurements provides significantly better error bars on the extracted parameters.

To assess those two points, we will simulate neutron-star gravitational wave and x-ray detections and estimate their associated errors:
\begin{itemize}
    \item A GW170817-like double neutron star binary merger emitting gravitational waves, for which individual masses and tidal deformabilities are recovered. One of the neutron stars of the binary has a mass ${M_{\rm gw}=1.5{\;\rm M}_{\odot}}$ to which the tidal deformability $\Lambda_{\rm gw}\simeq 470$ is determined by the unified equation of state model DD2 \citep{Typel2010} as presented in Ref.~\cite{Suleiman2021}. We use the precision reported in Ref.~\cite{Carson2019} as well as parameter estimation operated with the public software \textsc{Bilby} \citep{Ashton2019} to simulate the error associated to the mass and tidal deformability, along with a bivariate normal distribution peaked on $M_{\rm gw}$ and $\Lambda_{\rm gw}$. We use two detector sensitivities : projected O4 sensitivity of Laser Interferometer Gravitational Wave Observatory (LIGO) and Virgo facilities (denoted {GW-O4}) and third generation telescopes CE and ET sensitivity (denoted {GW-3G})\footnote{The sensitivity curves for CE and ET (ET-D design \cite{Hild2011}) used to simulate the signal in this paper can be found at \url{https://dcc.ligo.org/LIGO-T1500293/public}. Updated sensitivity ET-D curves were provided in Ref.~\cite{Branchesi-2023} recently, but conclusions of our paper remain robust.}.
    \item An x-ray source with a neutron star mass ${M_{\rm xray} = 1.4{\rm \;M}_{\odot}}$ and a radius established from the equation of state model DD2 $R_{\rm xray}\simeq 13.16$\;km. We use the precision reported in Ref.~\cite{riley2019} for the detector sensitivity denoted NICER and the projected sensitivity reported in Ref.~\cite{Ray2019} denoted STROBE-X, to simulate a distribution of masses and radii using a bivariate normal distribution peaked on M$_{\rm xray}$ and $R_{\rm xray}$. 
\end{itemize}

%-------------------------------------------------------------------------------------%
\subsubsection{Performing fits on agnostic equation of state sets}\label{sec:paramExtract/fits}

The usual approach for parameter extraction with quasiuniversal relations is to use parametrized functions (usually a polynomial) fitted to macroscopic parameters computed with a set of EoS, and reported with a precision that attests to the largest difference between the fit and the set's macroscopic parameters. 

The first fits available were based on a limited number of EoS models. For example, the fits presented in Ref.~\cite{Maselli13} used three EoSs with purely nucleonic cores; the reported precision of $2\%$ for the relation $C(\Lambda)$ was shown in Ref.~\cite{Suleiman2021} to be too small when compared to several EoSs of various stiffness. Recognizing that the so called "universal" relations were only quasi-independent of the description of neutron star's interior, efforts were made to fit to larger sets of nuclear physics based EoSs, including various core compositions thus increasing the reported error of the fits; for example, in Ref.~\cite{Yagi2017}, the fits were performed using 20 EoSs with nucleonic cores, seven EoSs with hyperonic or kaon cores and three quark stars; a reported precision of 6.5\% is given for neutron stars and 15\% for quark stars. As a neutron star's core composition remains unknown, a bias may be introduced by performing the fit on sets containing more nucleonic models than hybrid or hyperonic ones. Bias may also be introduced by the nuclear physics approach used for the set of EoSs on which the fits are based: relativistic mean field based models tend to be stiffer than, e.g., Skyrme energy density functional models. To mitigate the impact of these choices, we choose sets of EoSs that are more agnostic, as has been done, e.g., in Ref.~\cite{Godzieba2021,Legred2023, Pradhan2023}. 

In this paper, we fit each of the three relations presented in Sec.~\ref{sec:ur} with the different EoS sets considered in this paper. For each relation and each set, a nonlinear least square method is used to determine the parameter $a_k$ with $k\in[0,5]$ in
\begin{align} 
    C_{\rm fit}&=\sum_{k=0}^5a_k(\ln\Lambda)^k \;, \label{eq:cl} \\
    \ln \bar I_{\rm fit}&=\sum_{k=0}^5a_k \ln(\Lambda)^k \;, \label{eq:ic} \\
    \bar I_{\rm fit}&=\sum_{k=0}^5 a_k C^{-k} \label{eq:il} \;.
\end{align}

The fits are performed using macroscopic parameters of neutron stars with at least a mass of 1.0\;M$_{\odot}$ up to the last stable mass configuration. An error range is associated to each fit and denoted ${(\Delta X_{\rm fit})^{\rm S}}$ with $X$ the fitted quantity and $S$ the set of EoSs. The error is assessed from the full range of the (finite) 
nuclear set, and the 99\% percentiles of the GP+astro set and MM+$\chi$+PSR set. For the GP+astro and MM+$\chi$+PSR sets, there are no defined edge to the range because the EoS are drawn from an underlying distribution that can, in principle, extend to infinity. The fit parametrizations for $C(\Lambda)$, $\bar{I}(\Lambda)$, and $\bar{I}(C)$ presented in Ref.~\cite{Yagi2017} (later denoted Yagi \& Yunes) are also used as a comparison.

%-------------------------------------------------------------------------------------%
\subsubsection{Marginalization of the error}\label{sec:paramExtract/margi} 

To include the fit error ${(\Delta X_{\rm fit})^{\rm S}}$ in the extraction of a parameter, we specify a marginalization over this systematic uncertainty, as seen previously in e.g. Refs.~\cite{Abbott2018,Chatziioannou2018,Carson2019}. This is done by introducing an extra parameter $\delta X$ with a random distribution reflecting the fit uncertainty. Each extracted parameter is multiplied by a factor $(1+\delta X_{\rm fit}^S)$.  For example, when applying $C_\text{fit}$ as in Refs.~\cite{Abbott2018,Chatziioannou2018}, the marginalization consists of drawing the fit error parameter $\delta X$ from a Gaussian function peaked at zero with standard deviation ${(\Delta C_{\rm fit})^{\rm S}/3C_\text{fit}}$, so that results are within ${\Delta C_{\rm fit}}$ at 3-$\sigma$. We refer to this technique as the Gaussian marginalization. 

In the absence of information on the distribution of macroscopic parameters computed from the EoS set used to perform the fit, we propose that users choose a uniform distribution instead of a Gaussian one: ${\delta X_{\rm fit}^S \in [-(\Delta X_{\rm fit})^{\rm S};(\Delta X_{\rm fit})^{\rm S}]}$ is randomly drawn from a uniform distribution; we refer to this approach as uniform marginalization. It leads to a larger error on the extraction of the macroscopic parameter but avoids preferring the exact fit. To illustrate that the distribution of points is not necessarily a Gaussian for a given set of EoS, we present the distribution of compactness and dimensionless moment of inertia around the fit for equation of state sets in Appendix~\ref{sec:appendix,dis}.

We also show the extremes of the quasiuniversal relation as a way to represent the systematic uncertainty.
We consider ${X_{\rm fit \pm }^S(Z) = X_{\rm fit}^S(Z) \pm (\Delta X_{\rm fit})^{\rm S}}$. We perform the parameter extraction at each extreme of the range and obtain two distributions which represent the two extremes of the fit error. We refer to this approach as the fit limits.

%-------------------------------------------------------------------------------------%
%-------------------------------------------------------------------------------------%
% \subsection{Equation of state inference}

%-------------------------------------------------------------------------------------%
\begin{figure*}
\begin{subfigure}[b]{0.49\hsize}
\resizebox{\hsize}{!}{\includegraphics{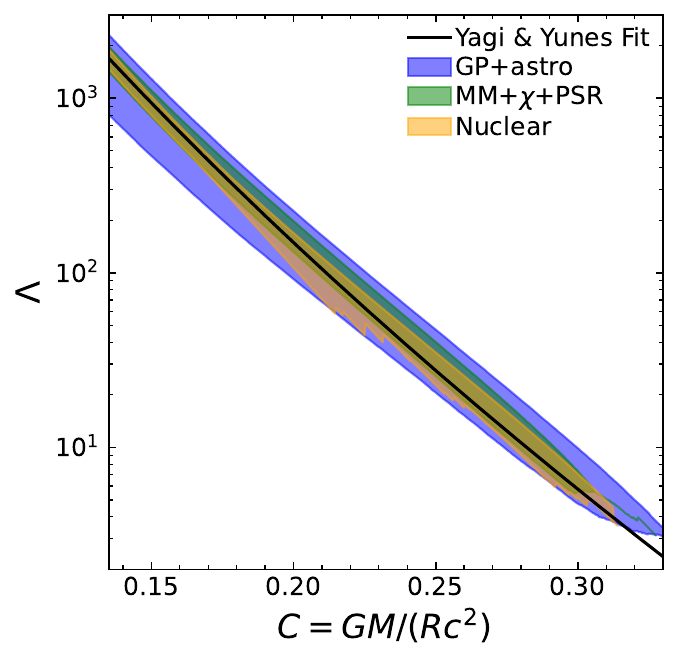}}
\caption{Relation between the compactness $C$ and the dimensionless tidal deformability $\Lambda$.}
\label{fig:LambdaC}
\end{subfigure}
\hfill
\begin{subfigure}[b]{0.49\hsize}
\resizebox{\hsize}{!}{\includegraphics{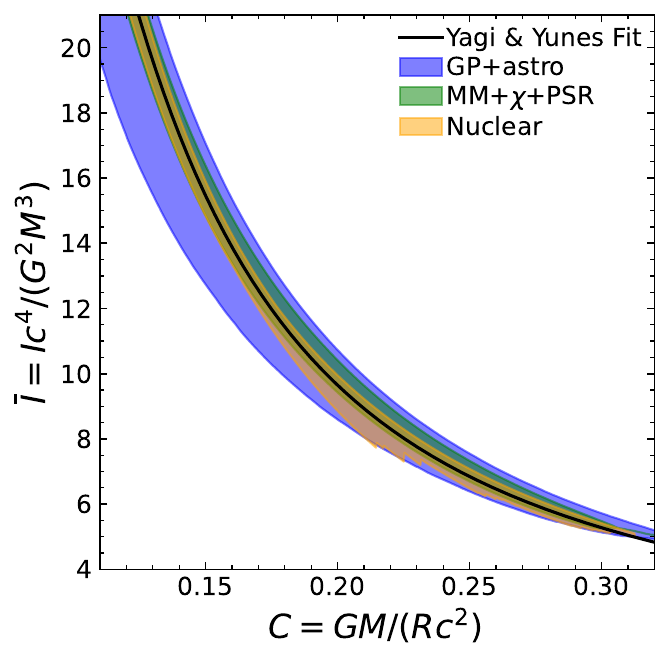}}
\caption{Relation between the compactness $C$ and the dimensionless moment of inertia $\bar{I}$.}
\label{fig:IbarC}
\end{subfigure}
\begin{subfigure}[b]{0.49\hsize}
\resizebox{\hsize}{!}{\includegraphics{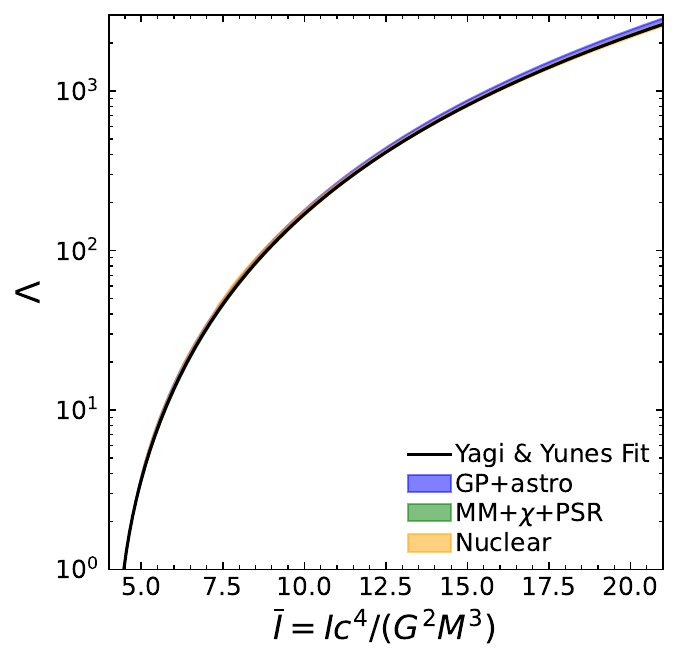}}
\caption{Relation between the dimensionless tidal deformability $\Lambda$ and the dimensionless moment of inertia $\bar{I}$.}
\label{fig:LambdaIbar}
\end{subfigure}
\caption{Contours (99 percentile) presented for the nuclear set, MM+$\chi$+PSR set and the GP+astro set. In black we present the Yagi \& Yunes fit \citep{Yagi2017}.}
\label{fig:urforsets}
\end{figure*}
%-------------------------------------------------------------------------------------%

%-------------------------------------------------------------------------------------%
\begin{figure*}
    \begin{subfigure}[b]{0.49\hsize}
        \resizebox{\hsize}{!}{\includegraphics{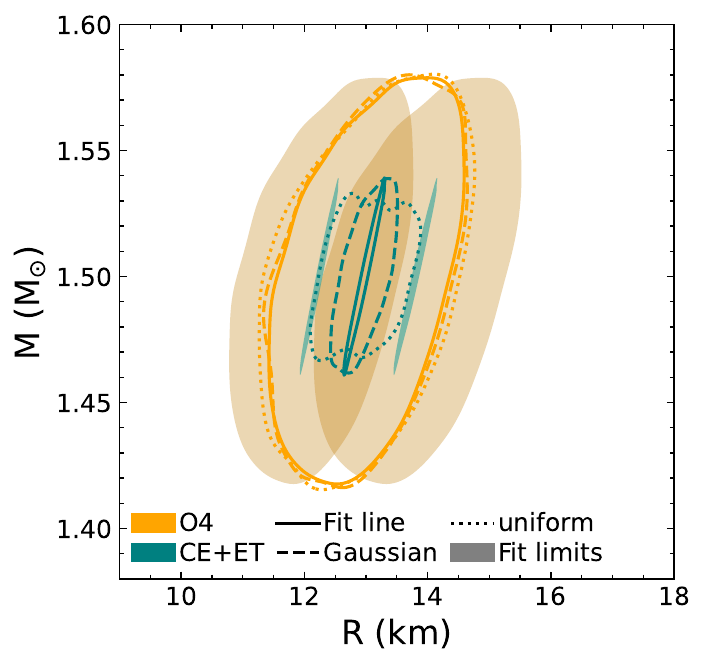}}
        \caption{MM+$\chi$+PSR set }
        \label{fig:mrgwmm}
    \end{subfigure}
    \hfill
    \begin{subfigure}[b]{0.49\hsize}
        \resizebox{\hsize}{!}{\includegraphics{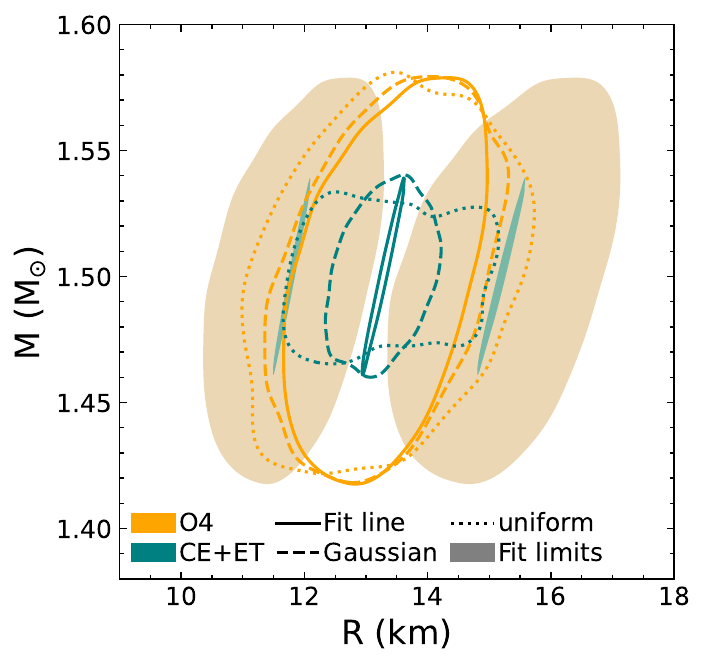}}
        \caption{GP+astro set}
        \label{fig:mrgwGP}
    \end{subfigure}
    \caption{$M(R)$ extracted from the fit of $C(\Lambda)$ and a simulated GW signal in the sensitivity of O4 (orange) and CE and ET (green); contours are shown at 1-$\sigma$. Results are presented for the exact fit (plain lines), with the Gaussian marginalization of the error (dashed line), and with the uniform marginalization of the error (dotted line); the limits of the error are presented in shades of colors.}
    \label{fig:mrgw}
\end{figure*}
%-------------------------------------------------------------------------------------%

%-------------------------------------------------------------------------------------%
%-----------------------------------  Sec. 3  -------------------------------------%
%-------------------------------------------------------------------------------------%
\section{Results}\label{sec:results}

%-------------------------------------------------------------------------------------%
%-------------------------------------------------------------------------------------%
\subsection{Comparison of the different equation of state sets} \label{res:compareSets}

In Fig.~\ref{fig:urforsets} we present the relations of Sec.~\ref{sec:ur} established from different sets of EoS: the nuclear set, the GP+astro and the MM+$\chi$+PSR set. In black is also presented the fits of Yagi \& Yunes. 

The agnostic GP+astro set presents the largest EoS variability. It explores a larger space than the nuclear set and the agnostic MM+$\chi$+PSR set, and overlaps them both, showing that the chiral effective field theory constraints used in the metamodeling approach offer stronger constraints than current astrophysical measurements from NICER and GW170817. We note also that the GP+astro set does not encompass the metamodel (nor the nuclear) set in a symmetric way, particularly at low compactness: the meta model set favors higher tidal deformabilities and moment of inertia, in other words, stiffer EoSs.  

The nuclear set and the meta model do not perfectly overlap, as some of the EoSs used in the nuclear set are not in accordance with chiral effective field theory constraints. The nuclear set is in good accordance with the fit of Yagi \& Yunes, which was also based on complete nuclear physics calculation EoSs. The nuclear set was established using 61 EoSs with various core compositions (including deconfined quarks), however it has stronger EoS invariance than the Yagi \& Yunes fit based on $\sim30$ EoSs. The error associated to the Yagi \& Yunes fit of $C(\Lambda)$ is of 6.5\% and 15\% excluding and including quark stars, respectively. Our fit for this relation gives a 6\% maximum error (see Table~\ref{tab:fitParam} in the Appendix~\ref{app:fitParam}) even though it includes a significant portion of hybrid models. The smaller error is related to the selection of modern EoSs which are calibrated to astrophysical and nuclear physics data: for example, the set used to establish the Yagi \& Yunes fit includes models with low radii at fixed mass (i.e., high tidal deformability and moment of inertia), for which microphysics parameters have been disfavored by nuclear physics laboratory experiments, e.g., WFF1 and WFF2 \cite{Wiringa1988}.

The relation $C(\Lambda)$ in Fig.~\ref{fig:LambdaC} and $\bar{I}(C)$ in Fig.~\ref{fig:IbarC} retain a clear EoS variability. The relation $\bar{I}(\Lambda)$ in Fig.~\ref{fig:LambdaIbar} is EoS invariant to the point that we cannot distinguish differences between the sets of EoSs: this relation is sufficiently universal to overcome the EoS variability emerging from the use of different EoS sets. This result is in accordance with Ref.~\cite{Legred2023} which quantifies the degree of universality of various relations: we also find subpercent errors on the fits (see Table~\ref{tab:fitParam} in the Appendix~\ref{app:fitParam}).

Finally, from this hierarchy of universality between the different relations, we can anticipate that there exists a detector sensitivity for which the radius extraction from the measurement of the tidal deformability in a gravitational wave signal could be overcome by the EoS variability, while the extraction of the moment of inertia from the same signal would not.

%-------------------------------------------------------------------------------------%
\subsection{Error marginalization}

In this section, we compare different error marginalization techniques for quasiuniversal relation applications as described in Sec.~\ref{sec:paramExtract/margi} in the context of current and future detections. For gravitational wave data (see Sec.~\ref{sec:UR/sources}), simulated $M$ and $\Lambda$ distributions are used in the fits of $C(\Lambda)$ and $\bar{I}(\Lambda)$, with different error marginalization, to obtain distributions of $R$ and $I$ respectively. For x-ray data (see Sec..~\ref{sec:UR/sources}), simulated $M$ and $R$ distributions are used in the fit of $\bar{I}(C)$, with different error marginalization, to obtain distributions of $I$.

%-------------------------------------------------------------------------------------%
\subsubsection{Radius from binary neutron star mergers} \label{res:margi,CL}

In Fig.~\ref{fig:mrgw}, we present the extraction of the radius from the simulated gravitational wave signal of a binary neutron star merger using the fit of the quasiuniversal relation $C(\Lambda)$; the O4 sensitivity is presented in orange and the CE and ET sensitivity is presented in green. Fig.~\ref{fig:mrgwmm} used the fit based on the metamodeling set of EoS while Fig.~\ref{fig:mrgwGP} used the Gaussian process set. 

The radius extraction using the Gaussian or the uniform marginalization of the error, or simply the fit line and without including any error, are (roughly) equivalent at O4 sensitivity. These fit limits (in shaded colors) are larger at that sensitivity but \text{comparable to or smaller than} to the size of the contour. 

In the case of third generation sensitivity, we first see that there is a significant difference with the fit line without inclusion of the error as well as between the Gaussian and uniform marginalization: the Gaussian marginalization has overestimated the precision on the radius by around a factor of 2 compared to uniform marginalization. We can conclude that the technique used to include the systematic error has an impact on the results at this sensitivity. We also see on that figure that the range of the radius extracted with the fit line is significantly smaller than after error marginalization: this indicates that the parameter extraction contours at that sensitivity are related to the fit error and not the the tidal deformability recovery from the gravitational wave signal. This is confirmed by the fit limits presented in shades of green: they are at the opposite ends of the uniform marginalization and do not overlap at all. 

Finally, the metamodel (MM+$\chi$+PSR) fit offers a better precision on the parameter extraction than the Gaussian process (GP+astro) fit, as the maximum error presented in Table~\ref{tab:fitParam} indicated. This indicates that in parameter extraction, current astrophysical constraints are less constraining than nuclear theory calculations. We also note that higher radii are extracted for a given mass with the Gaussian process set, in accordance with our discussion of Fig.~\ref{fig:LambdaC} in Sec.~\ref{res:compareSets}.

\begin{figure}
\resizebox{\hsize}{!}{\includegraphics{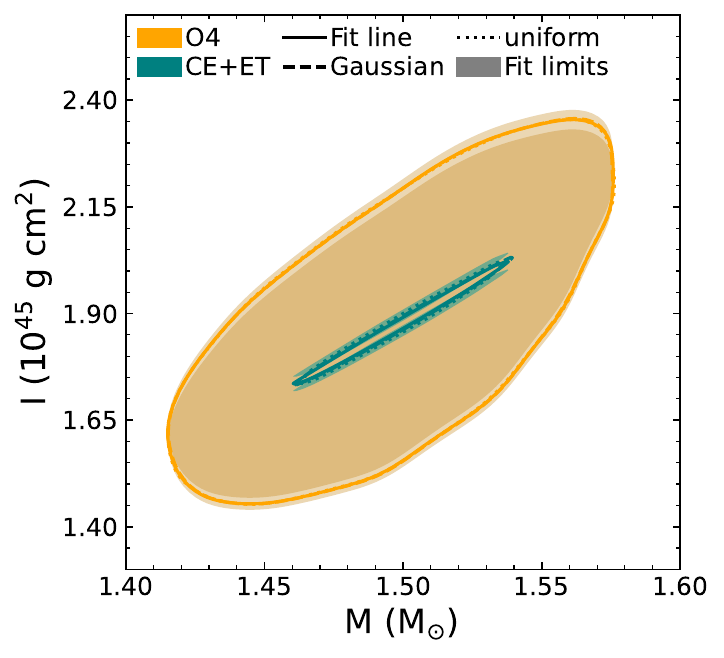}}
\caption{$I(M)$ extracted from the fit of $\bar{I}(\Lambda)$ and a simulated GW signal in the sensitivity of O4 (orange) and CE and ET (green); contours are shown at 1-$\sigma$. Results are presented for the fit line (plain lines), with the Gaussian marginalization of the error (dashed line), and with the uniform marginalization of the error (dotted line); the fit limits are presented in lighter shades. Fits used are that of GP+astro.}
\label{fig:imgwGP}
\end{figure}

%-------------------------------------------------------------------------------------%
\subsubsection{Moment of inertia from binary neutron star mergers}

\begin{figure*}
\begin{subfigure}[b]{0.49\hsize}
\resizebox{\hsize}{!}{\includegraphics{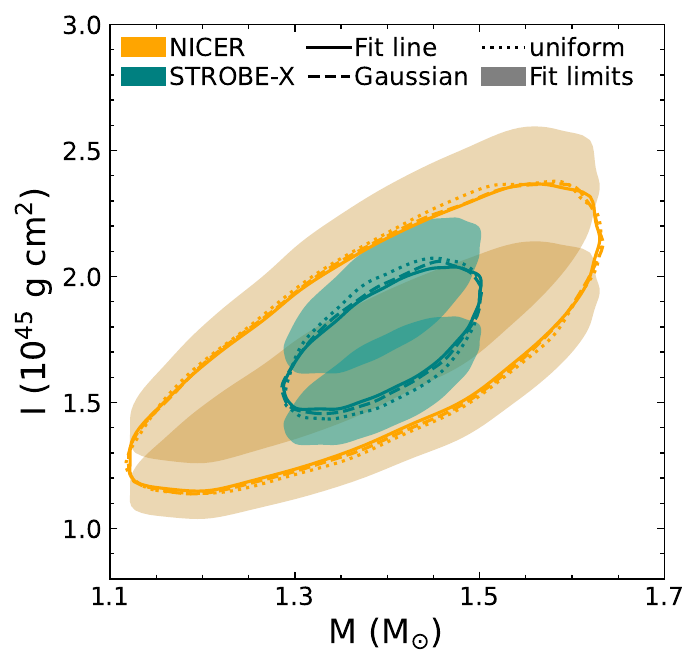}}
\caption{MM+$\chi$+PSR set}
\label{fig:imxraymm}
\end{subfigure}
\begin{subfigure}[b]{0.49\hsize}
\resizebox{\hsize}{!}{\includegraphics{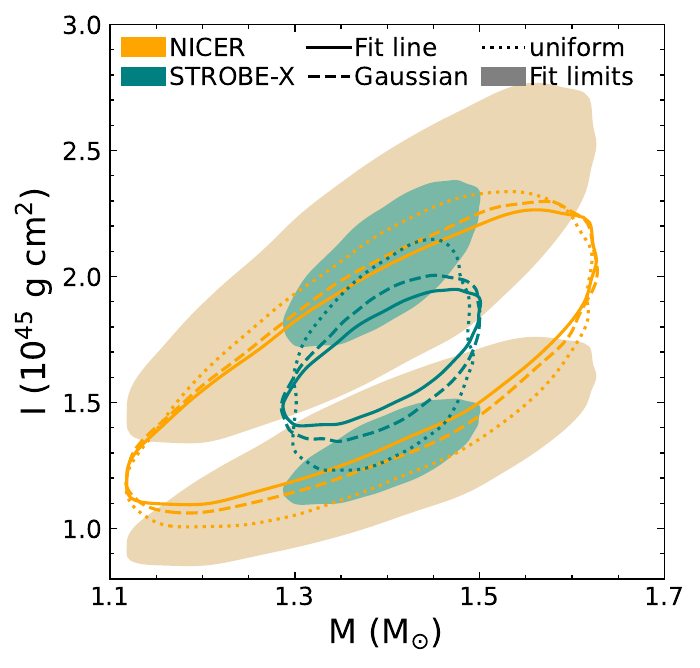}}
\caption{GP+astro set}
\label{fig:imxrayGP}
\end{subfigure}
\caption{$I(M)$ extracted from the $\bar{I}(C)$ relation and the simulated x-ray signal in the sensitivity of NICER (orange) and STROBE-X (green); contours are shown at 1-$\sigma$. Results are presented for the fit line (plain lines), with the Gaussian marginalization of the error (dashed line) and with the uniform marginalization of the error (dotted line); the fit limits are presented in lighter shades.}
\label{fig:imxray}
\end{figure*}

In Fig.~\ref{fig:imgwGP}, we present the extraction of the moment of inertia using the fit of $\bar{I}(\Lambda)$ based on the Gaussian process set (which produces the largest fit error, see Table~\ref{tab:fitParam}). In that case, we can see that the two marginalization techniques, the limit of the error and the extraction from the fit line all overlap, both for O4 sensitivity and CE+ET sensitivity. We use the same source and same set of EoS as for Fig.~\ref{fig:mrgw}, so this overlap results from the quasiuniversality of the relation $\bar{I}(\Lambda)$. As can be seen in Fig.~\ref{fig:LambdaIbar}, this relation has stronger universality, which makes considerations of the marginalization technique, or even purely of the error, irrelevant. The statistical uncertainty associated with the parameter estimation dominates whatever the sensitivity of the detector. We conclude that the universality of the relation $\bar{I}(\Lambda)$ will hold with the third generation of gravitational wave detectors.

%-------------------------------------------------------------------------------------%
\subsubsection{Moment of inertia from x-ray detections}

In Fig.~\ref{fig:imxray}, we present the extraction of the moment of inertia using the relation $\bar{I}(C)$. For the metamodel fits presented in Fig~\ref{fig:imxraymm}, the marginalization technique does not impact the results. For the fits based on the Gaussian process set presented in Fig.~\ref{fig:imxrayGP}, the results are impacted at high sensitivity (STROBE-X) as the Gaussian marginalization underestimates the error by almost a third compared to the uniform marginalization. The constraints brought forth by nuclear physics in the metamodel allow us to consider that STROBE-X sensitivity still dominates (or is at least equivalent) to the EoS variability of the metamodel based quasiuniversal relation $\bar{I}(C)$.

%-------------------------------------------------------------------------------------%
%-------------------------------------------------------------------------------------%
\subsection{Quasiuniversal relations vs EoS inference}

In this section, we compare the extraction of the radius and the moment of inertia using two methods: the quasiuniversal fits and EoS inference. We consider a simulated CE+ET gravitational wave detection recovering $M$ and $\Lambda$ and a simulated STROBE-X detection recovering $M$ and $R$, following the description of Sec.~\ref{sec:UR/sources}. For quasiuniversal relations, we use the distributions of $R$ and $I$ presented in Fig.~\ref{fig:mrgw}, Fig.~\ref{fig:imgwGP}, and Fig.~\ref{fig:imxray} in the case of uniform marginalization.

For EoS inference, we use the simulated data to infer constraints on the EoS with a simple likelihood estimation on the 1-$\sigma$ data and the MM+$\chi$+PSR and GP+astro sets. We then use the updated EoS distribution and general relativity to compute $R$ and $I$ from the mass distributions of the of the sources using the methods of Sec.~\ref{sec:methods}.

%-------------------------------------------------------------------------------------%
\subsubsection{Moment of inertia extraction}

%-------------------------------------------------------------------------------------%
\begin{figure*}
\begin{subfigure}[b]{0.49\hsize}
\resizebox{\hsize}{!}{\includegraphics{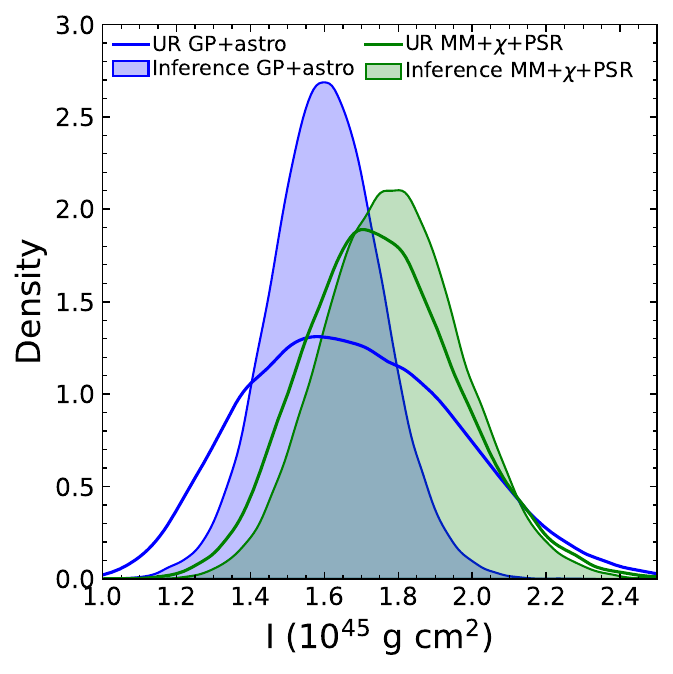}}
\caption{From STROBE-X x-ray data.}
\label{fig:disxray}
\end{subfigure}
\hfill
\begin{subfigure}[b]{0.49\hsize}
\resizebox{\hsize}{!}{\includegraphics{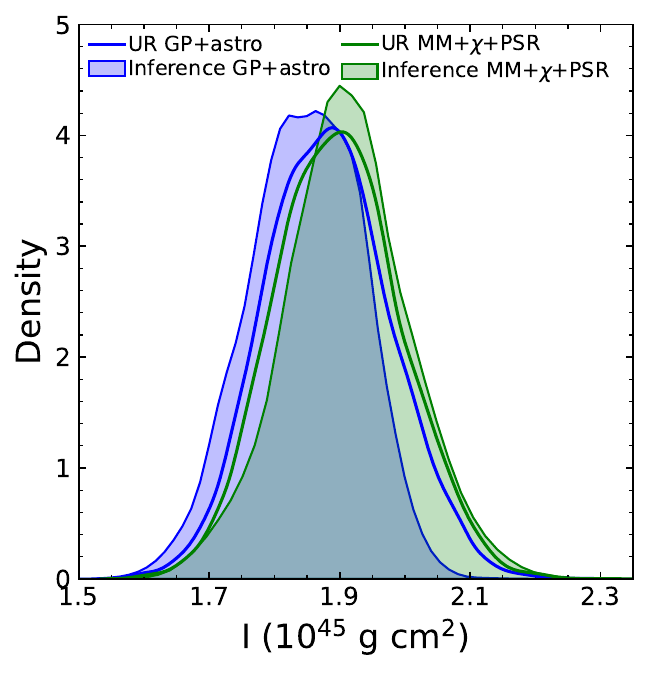}}
\caption{From CE+ET gravitational wave data.}
\label{fig:disgw1}
\end{subfigure}
\caption{Moment of inertia distribution from parameter extraction operated with quasiuniversal relation fits and EoS inference.}
\label{fig:distrib}
\end{figure*}
%-------------------------------------------------------------------------------------%

We show the moment of inertia distribution extracted  from the simulated STROBE-X source in Fig.~\ref{fig:disxray}. We perform EoS inference with either the GP+astro and MM+$\chi$+PSR EoS sets as priors, and compute the resulting $I$ distribution using the updated EoSs and the observed masses. We compare this to the moment of inertia distribution obtained with the quasiuniversal relation fits which parameters are presented in Table~\ref{tab:fitParam}. The EoS inference leads to a better determination of the moment of inertia than the use of quasiuniversal relation fits. In the case of the GP+astro set, the improvement is significant, while in the case of the metamodel it is almost equivalent: this is in accordance with results presented in Fig~\ref{fig:imxray}. The peaks of the EoS inference distributions depend on the EoS prior and are not the same for the GP+astro and the metamodel. The GP+astro set is informed by GW170817, which softens the EoS distribution, while the metamodel set allows stiffer models. The distribution for $I$ determined by the quasiuniversal relation fits with the GP+astro extracts higher values of the moment of inertia than its EoS inference counterpart. This is because the quasiuniversal fits discard some of the prior information encoded in the full EoS set; it considers only the relation between compactness and $I$ rather than the preferred values of compactness within that relation, extracting values linked only to the current observation.

The moment of inertia distribution extracted with gravitational wave data simulated with CE and ET sensitivity is presented in Fig.~\ref{fig:disgw1}. The extraction of $I$ with the quasiuniversal relation fits of $\bar{I}(\Lambda)$ are very similar: as discussed in Fig.~\ref{fig:LambdaIbar}, this relation is very universal. The difference of the peak's values in the distributions for EoS inference and quasiuniversal relation fits is also visible even though the distributions overlap well; both EoS sets generate similar fit results, as those discard prior information preferring softer or stiffer EoS. Finally, we also note that the 3G gravitational wave detection offer a better constraint on the moment of inertia than STROBE-X does with EoS inference.

%-------------------------------------------------------------------------------------%
\subsubsection{Radius extraction}
%-------------------------------------------------------------------------------------%
\begin{figure}
\resizebox{\hsize}{!}{\includegraphics{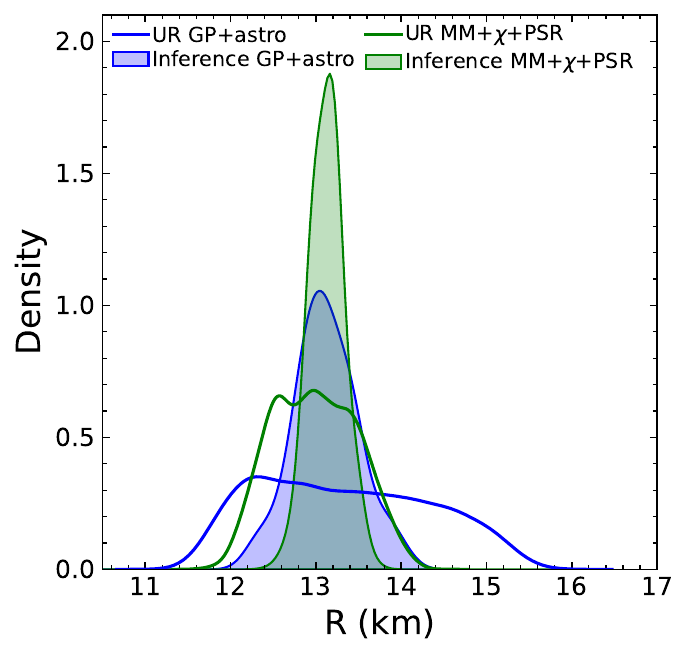}}
\caption{Radius distribution from parameter extraction for simulated CE+ET gravitational wave data, using quasiuniversal relation fits and EoS inference.}
\label{fig:disgw2}
\end{figure}
%-------------------------------------------------------------------------------------%
We show the radius distribution extracted with the fit of the quasiuniversal relation fit of $C(\Lambda)$ from the GP+astro and MM+$\chi$+PSR EoSs, and from the updated EoS distributions inferred from the simulated gravitational wave data source in Fig.~\ref{fig:disgw2}. Results are similar to Fig.~\ref{fig:disgw1}, except that the distribution offered by quasiuniversal relation fit is much larger. The extraction of the radius from the GP+astro set leads to a range extending to very large radii. The systematic bias due to the use of fit in the extraction of the radius was also discussed in Ref.~\cite{Kashyap2022,Huxford2023}. We note that the metamodel inference peak on the radius in Fig~\ref{fig:disgw2} is much narrower than the peak in moment of inertia of Fig.~\ref{fig:disgw1}.  This is because the radius is strongly related to the description of the crust while the moment of inertia is mainly sensitive to the core. While the core treatment of the MM+$\chi$+PSR and the GP+astro seems to lead to similar distributions for the moment of inertia inference, this is not the case for the radius inference. The power of the low density treatment of the meta model explicitly shows in this case. This shows the impact of the nuclear physics information in the MM+$\chi$+PSR, which limits the EoS range as seen in Fig.~\ref{fig:setPnb}.

We note that the distributions presented in Fig.~\ref{fig:distrib} and Fig.~\ref{fig:disgw2} are not always very well resolved. In this paper, we have used precomputed EoS sets as priors for inference following methods described in Refs.~\cite{Landry2018,Essick2020, Legred2021}. This will be an issue for future detectors; a generative model for additional draws from the EoS prior operating with the inference would be necessary to offer well-resolved distributions.

We also note that for this work we have considered a single measurement that offers one constraint on the extracted parameter. If multiple measured parameters allow extraction with more than one technique, then it would be possible to find tension in the extracted values, as discussed in Ref.~\cite{Lioutas2021}. Such tension indicates that the prior assumptions of the EoS set have been violated, or in the case of a quasiuniversal extraction that the true EoS has a relation outside the fit error, and could suggest currently-unmodeled features in the EoS.

% %-------------------------------------------------------------------------------------%
% %-----------------------------------  Sec. 4  -------------------------------------%
% %-------------------------------------------------------------------------------------
\section{Conclusion}

In this paper, we discuss three macroscopic quasiuniversal relations : $C(\Lambda)$, $\bar{I}(C)$, and $\bar{I}(\Lambda)$. We have studied the use of these quasiuniversal relations in moment of inertia and radius extraction from gravitational wave and x-ray signals with current and future detector sensitivity.

We have used three different sets of EoSs to calibrate the relations and their associated error distributions. We have shown that for $C(\Lambda)$ and $\bar{I}(C)$ relations, the variability of EoS in these sets leads to differences in their corresponding quasiuniversal fit, quantified here by a larger fit error, while the $\bar{I}(\Lambda)$ presents a more pronounced universality. 
% To accurately express the EoS set dependence of the relations, we refer to data-driven  quasiuniversal relations instead of quasiuniversal relations. 

We have discussed the fit error marginalization in parameter extraction and have shown that different approaches do not significantly impact the result for O4 or NICER sensitivity, but do impact the result for Einstein Telescope and Cosmic Explorer and STROBE-X sensitivity. 

Finally, we have shown that using quasiuniversal relation fits in parameter extraction with the future detector sensitivity will overestimate uncertainties on the extracted parameter when compared to direct EoS inference. In general, when the quasiuniversality fit error gives larger variation in the extracted parameter than the uncertainty on that parameter from the measurement error, using quasiuniversal relation fits no longer reflects the information gained about that parameter from the observation and its EoS implications.

At current observational precision, quasiuniversal relations give broadly equivalent results for parameter extraction when compared to direct EoS inference for the same quantities. We show here that quasiuniversal relations become less effective when projecting results for future observations, even when conditioned on EoS knowledge from current precision measurement. Instead, the direct EoS constraints inferred from future observations should be included when determining observational implications to reveal the full capability of next-generation facilities.

\begin{acknowledgments}

The authors thank Rory Smith, Sunny Ng, Philippe Landry, Carl-Johan Haster, Isaac Legred and Emily Wuchner for useful discussion about numerical applications. The authors thank collaborators of the Nuclear Physics from multimessenger Mergers (NP3M) Focused Research Hub for useful discussions. The authors acknowledge the financial support of the National Science Foundation Grants No. PHY 21-16686 and No. PHY 21-10441.  The authors are grateful for computational resources provided by the LIGO Laboratory and supported by National Science Foundation Grants No. PHY-0757058 and No. PHY-0823459. Software: This work makes use of scipy \cite{Virtanen2020}, numpy \cite{harris2020} and matplotlib \cite{Hunter2007}.

\end{acknowledgments}

\appendix

\section{Equation of state relation for agnostic sets}\label{app:EoSSets}
\begin{figure}
    \centering
    \resizebox{\hsize}{!}{\includegraphics{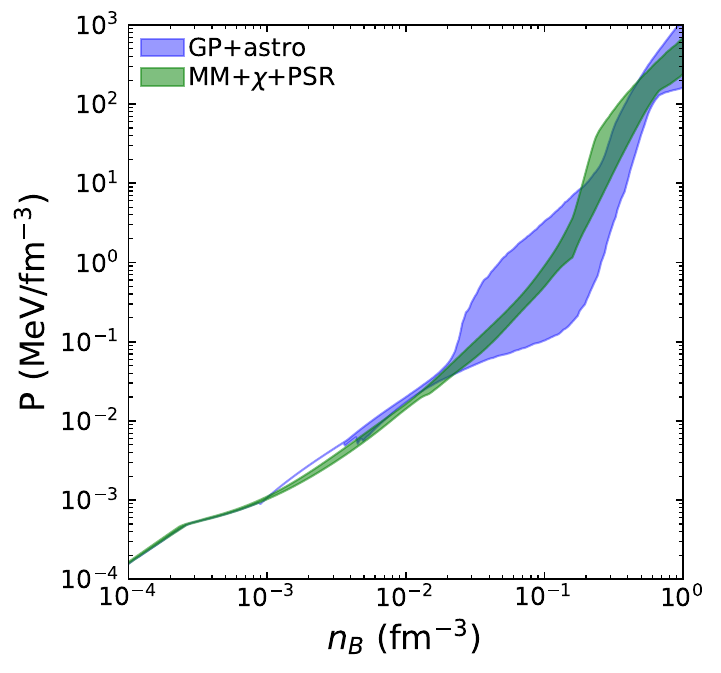}}
    \caption{Contours (99 percentile) for the relation between the pressure and the baryonic density for the GP+astro and MM+$\chi$+PSR agnostic sets.}
    \label{fig:setPnb}
\end{figure}

Figure~\ref{fig:setPnb} presents the equation of state (relation between the pressure $P$ and the baryonic density $n_B$) for the two agnostic sets discussed in this paper. 

\section{Mass, radius, moment of inertia and tidal deformability modeling} \label{sec:appendix,modeling}

We study the astrophysical features of a nonrotating neutron star within the gravity theory of general relativity, assuming the line element ${\rm d}s$ of a spherically symmetric, static and isotropic space-time determined by the metric $g_{\mu \nu}$ as
\begin{align}
    {\rm d}s^2 &= g_{\mu \nu} {\rm d}x^{\mu } {\rm d}x^{\nu}  \\
    &= -e^{2\phi(r)} {\rm d}t^2 + e^{2\lambda(r)}{\rm d}r^2 + r^2 {\rm d}\theta + r^2 \sin(\theta) {\rm d}\phi^2 \;, \nonumber
\end{align}
with $r$ the Schwarzschild-like radial coordinate. Considering a vacuum outside the neutron star, this metric reduces to the Schwarzschild one. Inside the star, the functions $\phi$ (gravitational redshift) and $\lambda$ (radial gravitational distortion) are solved using hydrostatic equilibrium equations and the EoS ${P = P(\epsilon)}$, with $P$ the pressure and $\epsilon$ the energy density. In the assumption that the matter inside the neutron star is a perfect fluid, we obtain the TOV differential equations 
\begin{align}
    &\frac{{\rm d}m}{{\rm d}r} =  \frac{4 \pi r^2}{c^2} \epsilon(r) \;, \\
    &\frac{{\rm d}\phi}{{\rm d}r} = \frac{Gm(r)}{r^2}  \bigg(1+ \frac{4\pi r^3 P(r)}{m(r)c^2} \bigg) \bigg( 1-\frac{2Gm(r)}{rc^2} \bigg)^{-1} \;, \\
    &\frac{{\rm d}P}{{\rm d}r} = -  \bigg( \frac{\epsilon(r)+P(r)}{c^2} \bigg) \frac{{\rm d}\phi}{{\rm d}r}  \;,
\end{align}
with $G$ the gravitational constant, $c$ the speed of light, and $m(r)$ the gravitational mass radial profile. The central energy density or central pressure, used as a boundary condition to solve the TOV equations, determines the total gravitational mass $M$ and the total radius $R$ of the neutron star. 

The dimensionless tidal deformability expresses the quadruple moment response induced on the neutron star by an external gravitational field. Its value at the surface of the star, denoted $\Lambda = \lambda_2(R)$, is given by the tidal deformability radial profile,
\begin{equation}
  \lambda_2(r) = \frac{2}{3} k_2(r)C(r)^{-5}\, , 
\label{eq:lambda} 
\end{equation} 
with $C(r) = Gm(r)/(rc^2)$ the compactness of a star of radius $r$. The tidal Love number $k_2$ is given by 
\begin{align}
    k_2(r) =& \,\frac{8C(r)^5}{5}\big(1-2C(r)\big)^2\big[2+2C(r)\big(y(r)-1\big)-y(r)\big] \nonumber \\
    & \times \bigg( 2C(r)\big[6-3y(r)+3C(r)\big(5y(r)-8\big)\big] \nonumber \\ 
    &+ 4C(r)^3\big[13-11y(r)+C(r)\big(3y(r)-2\big) \nonumber \\
    & \hspace{1.5cm} +2C(r)^2\big(1+y(r)\big)\big] \nonumber \\ 
    &+ 3\big(1-2C(r)\big)^2\big[2-y(r)+2C(r)\big( y(r)-1 \big) \big] \nonumber \\
    & \hspace{2.5cm} \ln \big(1-2C(r) \big) \bigg)^{-1} \;, 
    \label{eq:k2}
\end{align}
see, e.g., Ref.~\cite{Hinderer2008} and reference therein. The function $y(r)$ is to be solved simultaneously with the TOV equations using the additional differential equation \citep{Hinderer2008, Postnikov2010}  
\begin{equation}\label{y_eq}
    r \dfrac{ {\rm d} y}{{\rm d} r}+ y(r)^2 + F(r) y(r) + Q(r)=0 \, ,
\end{equation}
with the boundary condition $y(0)=2$, see Sec.~IV.A of Ref.~\cite{Damour2009}. The functions $F(r)$ and $Q(r)$ are given by
\begin{align}
    F(r)=& \bigg(1-\frac{4\pi G}{c^2} r^2\frac{\epsilon(r)-P(r)}{c^2} \bigg) \bigg( 1-\frac{2Gm(r)}{r c^2}\bigg)^{-1}\, , \label{F(r)}\\
    Q(r)=&\frac{4 \pi G }{c^2} r^2 \bigg( 1-\frac{2Gm(r)}{r c^2}\bigg)^{-1}\Biggl[\frac{5\epsilon(r)+9P(r)}{c^2} \nonumber \\
    &\hspace{0.3cm}+\frac{\epsilon(r)+P(r)}{c_s(r)^2}-\frac{6\,  c^2}{4\pi  Gr^2}\Biggr]  \; \label{Q(r)} \\
    & \hspace{0.3cm} -4\bigg( \frac{Gm(r)}{r c^2}+\frac{4\pi G}{c^4}r^2 P(r)\bigg)^2 \bigg( 1-\frac{2Gm(r)}{r c^2}\bigg)^{-2} \nonumber
\end{align}
with $c_s$ the sound speed that should be treated with caution around a discontinuous density, see, e.g., Ref.~\cite{Takatsy2020}. 

To model the moment of inertia, we use the slow and rigid rotation approximation detailed in Ref.~\cite{Hartle1967}\footnote{Modeling the moment of inertia outside of this approximation requires solving the Einstein's equations with a metric describing a stationary and axisymmetric space-time, see, e.g., the \href{http://www.lorene.obspm.fr}{LORENE} library.}. When considering rigid rotation, the uniform angular frequency is contributed to by the local spin frequency $\omega(r)$ and the angular momentum $j(r)$ of a sphere of radius $r$. We denote $\Omega$, the uniform angular frequency of the star, defined at the surface (${r=R}$) as  
\begin{equation}
    \Omega=\omega(R)+\frac{2Gj(R)}{c^2R^3} \;.
\end{equation}
The uniform angular frequency, the angular momentum and the moment of inertia are solutions of the differential equations,
\begin{align}
    & \frac{{\rm d}I}{{\rm d}r} =\frac{8\pi}{3}\, \frac{r^4}{e^{\phi(r)}}\frac{\omega(r)}{\Omega}\frac{\epsilon(r)+P(r)}{c^2}\bigg( 1-\frac{2Gm(r)}{c^2 r} \bigg)^{-1/2}\;, \label{eq:tov1}\\
    & \frac{{\rm d}\omega (r)}{{\rm d}r}=\frac{6 G }{c^2 } \frac{e^{\Phi(r)}}{r^4} j(r) \bigg(1-\frac{2Gm(r)}{c^2 r} \bigg) ^{-1/2}\;,  \label{eq:tov2} \\
    &\frac{{\rm d}j(r)}{{\rm d}r} = \frac{8\pi }{3} \frac{r^4}{e^{\phi(r)}} \omega(r) \frac{\epsilon(r)+P(r)}{c^2}\bigg(1-\frac{2Gm(r)}{c^2 r} \bigg)^{-1/2}  \label{eq:tov3} \;,
\end{align}
to be solved simultaneously with the TOV equations.

%-------------------------------------------------------------------------------------%
%-------------------------------------------------------------------------------------%
%-------------------------------------------------------------------------------------%
\section{Distribution of EoSs in the sets}\label{sec:appendix,dis}

\begin{figure*}
    \resizebox{\hsize}{!}{\includegraphics{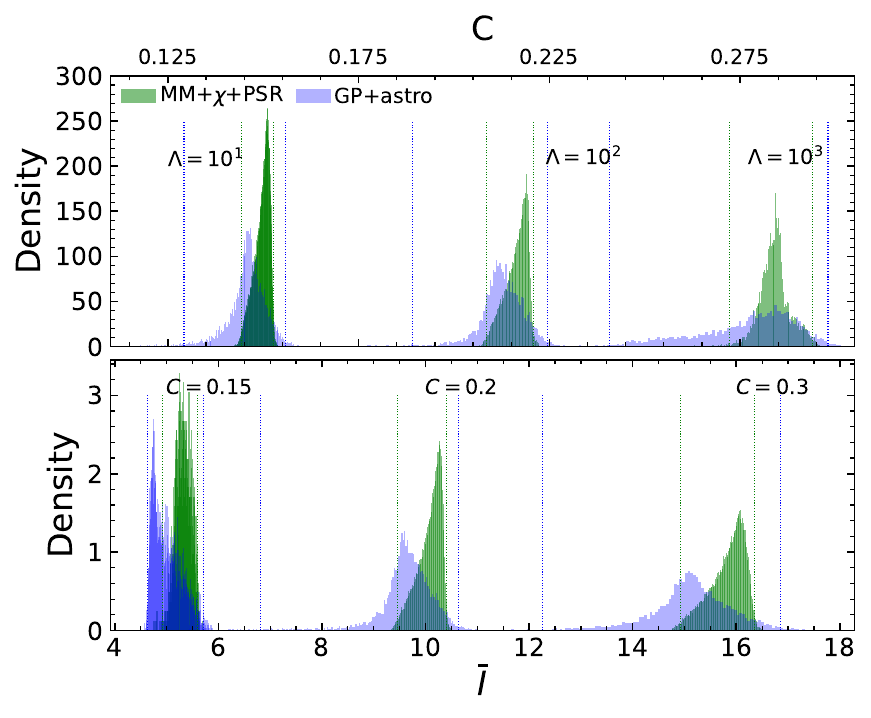}}
    \caption{Distribution of points of compactness (top) and dimensionless moment of inertia (bottom) for a few values of the tidal deformability and compactness respectively, in MM+$\chi$+PSR and GP+astro sets. Vertical lines correspond to the 99 percentile limits of the distributions.}
    \label{fig:disset}
\end{figure*}

We present in Fig.~\ref{fig:disset} the distribution of compactness and dimensionless moment of inertia points for GP+atro and MM+$\chi$+PSR set. 

%-------------------------------------------------------------------------------------%
%-------------------------------------------------------------------------------------%
%-------------------------------------------------------------------------------------%
\section{Fit parameters and errors associated to the fits}\label{app:fitParam}

We present in Table~\ref{tab:fitParam} the parameters of the fit for the relations presented in Sec.~\ref{sec:paramExtract/fits} and the maximum error associated to the fits with the different EoS sets presented in this paper. 

\begin{table*}
    \centering
    \begin{tabular}{|c||c||cccccc|c|}
    \hline
       Relation & Set & $a_0$ & $a_1$ & $a_2$ & $a_3$ & $a_4$ & $a_5$ & $\Delta_{\rm max}$ (in \%)\\ \hline \hline
       \multirow{3}{*}{$C(\Lambda)$} & GP+astro &	3.7678 $\times 10^{-1}$ & 	-5.1851 $\times 10^{-2}$ & 	8.3659 $\times 10^{-3}$ & 	-1.6529 $\times 10^{-3}$ & 	1.5470 $\times 10^{-4}$ & 	-5.0440 $\times 10^{-6}$ & 	12.6129 \\ 
         & Nuclear &	3.5636 $\times 10^{-1}$ & 	-3.6950 $\times 10^{-2}$ & 	4.5372 $\times 10^{-3}$ & 	-1.1316 $\times 10^{-3}$ & 	1.1773 $\times 10^{-4}$ & 	-4.0028 $\times 10^{-6}$ & 	5.9967 \\ 
         & MM+$\chi$+PSR &	3.3888 $\times 10^{-1}$ & 	-1.0262 $\times 10^{-2}$ & 	-9.8856 $\times 10^{-3}$ & 	2.0656 $\times 10^{-3}$ & 	-1.9511 $\times 10^{-4}$ & 	7.2814 $\times 10^{-6}$ & 	5.2443 \\ 
        \hline \hline
        \multirow{3}{*}{$\bar{I}(\Lambda)$} & GP+astro &	1.5011 $\times 10^{0}$ & 	5.2134 $\times 10^{-2}$ & 	2.4508 $\times 10^{-2}$ & 	-9.9877 $\times 10^{-4}$ & 	2.7064 $\times 10^{-5}$ & 	-3.3703 $\times 10^{-7}$ & 	0.9699 \\ 
         & Nuclear &	1.4874 $\times 10^{0}$ & 	6.9698 $\times 10^{-2}$ & 	1.9206 $\times 10^{-2}$ & 	-2.5145 $\times 10^{-4}$ & 	-2.0918 $\times 10^{-5}$ & 	7.4502 $\times 10^{-7}$ & 	0.6774 \\ 
         & MM+$\chi$+PSR &	1.4857 $\times 10^{0}$ & 	6.7440 $\times 10^{-2}$ & 	1.8129 $\times 10^{-2}$ & 	2.2972 $\times 10^{-4}$ & 	-7.7256 $\times 10^{-5}$ & 	2.8044 $\times 10^{-6}$ & 	0.5033 \\ 
        \hline \hline
        \multirow{3}{*}{$\bar{I}(C)$}  & GP+astro &	1.5333 $\times 10^{1}$ & 	-1.1303 $\times 10^{1}$ & 	3.8536 $\times 10^{0}$ & 	-5.0621 $\times 10^{-1}$ & 	3.2323 $\times 10^{-2}$ & 	-7.5871 $\times 10^{-4}$ & 	22.2608 \\ 
         & Nuclear &	6.1602 $\times 10^{0}$ & 	-4.7746 $\times 10^{0}$ & 	2.1004 $\times 10^{0}$ & 	-2.8280 $\times 10^{-1}$ & 	1.9466 $\times 10^{-2}$ & 	-4.9244 $\times 10^{-4}$ & 	9.6337 \\ 
         & MM+$\chi$+PSR &	2.3584 $\times 10^{0}$ & 	1.9423 $\times 10^{-1}$ & 	-4.5225 $\times 10^{-2}$ & 	1.1677 $\times 10^{-1}$ & 	-1.4423 $\times 10^{-2}$ & 	5.7714 $\times 10^{-4}$ & 	9.1280 \\ 
        \hline
    \end{tabular}
    \caption{Fit parameters of the relation $C(\Lambda)$, $\bar{I}(C)$, and $\bar{I}(\Lambda)$ based on the parametrization presented in Eqs.~\eqref{eq:cl}, \eqref{eq:ic}, \eqref{eq:il}, and maximum error associated to the fit $\Delta_{\rm max}$. Results are presented for the EoS sets discussed in this paper, GP+astro, the nuclear set and the MM+$\chi$+PSR.}
    \label{tab:fitParam}
\end{table*}

% \bibliography{biblio}
%merlin.mbs apsrev4-1.bst 2010-07-25 4.21a (PWD, AO, DPC) hacked
%Control: key (0)
%Control: author (72) initials jnrlst
%Control: editor formatted (1) identically to author
%Control: production of article title (-1) disabled
%Control: page (0) single
%Control: year (1) truncated
%Control: production of eprint (0) enabled
%

\end{document}